\documentclass[aps, prl, twocolumn,showpacs, showkeys, preprintnumbers, nofootinbib, superscriptaddress]{revtex4-2}
\usepackage[sort&compress]{natbib}
\usepackage{multirow}
\usepackage{graphicx}
\usepackage{amssymb,amsbsy,amsmath,amsfonts,longtable}
\usepackage{slashed}
\usepackage{soul}
\usepackage[latin1]{inputenc}

\begin{document}

\title{A glueball puzzle explained in terms of $\phi h_1(1380)$ dynamics: Aka $X(2370)$}
\author{K. P. Khemchandani}
\email{kanchan.khemchandani@unifesp.br}
\affiliation{Universidade Federal de Sao Paulo, C.P. 01302-907, Sao Paulo, Brazil.}

\author{A. Mart\'inez Torres}
\email{amartine@if.usp.br}
\affiliation{ Instituto de F\'isica, Universidade de S\~ao Paulo, C.P. 66318, 05389-970 S\~ao 
Paulo, SP, Brazil.
}

\author{E. Oset}
\email{oset@ific.uv.es}
\affiliation{Departamento de F\'isica Te\'orica and IFIC, Centro Mixto Universidad de Valencia-CSIC,
Institutos de Investigaci\'on de Paterna, Aptdo. 22085, E-46071 Valencia, Spain.}

\begin{abstract}
We study the interaction of $\phi$ with $h_1(1380)$, considering the latter as a $K^*\bar K$ molecular state, and find the generation of a state with the same quantum numbers, together with a mass and width compatible with the experimental results, as those of $X(2370)$, recently advocated as a candidate for a glueball. Within our description, the suppressed decay modes of $X(2370)$ to $\gamma \omega$, $\omega \phi$ and $K^*\bar K$, reported in the experimental paper as clues to its glueball nature, are naturally explained. Conversely, the non $K^* \bar K$ decay to $K \bar K \pi$ (the latter being the invariant mass where the resonance is observed) is shown to be largely enhanced due to the combination of a triangle singularity and the presence of $K(1630)$ in the $\rho \phi \to \bar K$ vertex involved in the triangle loop. 
\end{abstract}

\maketitle
\section{Introduction} 
The BESIII collaboration has studied the decay properties of $X(2370)$ ($J^{PC}=0^{-+}$) from which a glueball nature for the state has been claimed~\cite{BESIII:2026mvn}. Specifically, the decay $X(2370)\to K^0_SK^0_S\pi^0$ is analyzed, and the collaboration observes that this process does not proceed via $X(2370)\to K^{*0}(892) \bar K^0$. This, together with an observed suppression of the radiative decays of the state to $\omega$ and $\phi$, is interpreted by the collaboration as evidence that $X(2370)$ is a flavor singlet and qualifies as a glueball. These claims have been supported by a work using a glueball-quark Lagrangian, bosonizing the quark current into a series of three pseudoscalar mesons~\cite{Wang:2026iyq}. A different study~\cite{Lu:2026foy} concludes that a flavor-singlet radial hybrid is compatible with current data but remains indistinguishable from a mixed-glueball interpretation. In Ref.~\cite{Zhu:2026zju}, it is suggested to measure $\psi(2S) \to X(2370) + \gamma$ in addition to $J/\psi \to X(2370) + \gamma$, since the ratio of these decay widths is highly sensitive to the glueball-charmonium mixing angle. On the other hand, in Ref.~\cite{Ben:2026afy}, the authors allege that the arguments used in Ref.~\cite{BESIII:2026mvn} to claim the glueball nature of $X(2370)$ are flawed and the conclusion unjustified, and that all observed properties of $X(2370)$ can be naturally explained by considering the former to be a $\Sigma \bar \Sigma$ molecular state. 

We wish to contribute to the debate by presenting a different description, which, with no free parameters, reproduces the mass and width of $X(2370)$. At the same time, the suppression of the decay modes used to claim the glueball nature of the state comes naturally in this picture, and the $K^0 \bar K^0 \pi^0$ decay mode, different from the $K^{*0} \bar K^0$ one, is enhanced by a triangle singularity.  All these mentioned properties, as we will show, can be explained by considering  $X(2370)$ as a molecular state of $\phi h_1(1380)$, with  $h_1(1380)$ being itself a molecular state made mostly from the $K^* \bar K$ interaction. Since $\phi$ has quark content $s\bar s$ and $X(2370)$ has a remarkable affinity to decay to final states involving strangeness, such as $K^0_S K^0_S\pi^0$, multi-hadron systems provide a powerful window to understand its internal dynamics without invoking pure gluonic nature. Actually, the relevance of such systems has already been stressed in Ref.~\cite{Dong:2022cuw}, suggesting that, for energies at about 2.3 GeV, six-quark configurations may become increasingly important. 
 
 \section{Formalism}
The quantum numbers of $X(2370)$ are $J^{PC}=0^{-+}$ and its isospin is $I=0$. These quantum numbers are naturally obtained from the interaction of the vector meson $\phi$, $I^G(J^{PC})=0^- (1^{--})$, and the axial meson $h_1(1380)$, $I^G(J^{PC})=0^- (1^{+-})$. According to Refs.~\cite{Lutz:2003fm,Roca:2005nm}, the properties of $h_1(1380)$\footnote{Although the nominal mass reported for this state by the Particle Data Group (PDG) is $1409^{+9}_{-8}$~\cite{ParticleDataGroup:2026mpi}, the latest determination from the BESIII collaboration~\cite{BESIII:2022zel} gives the value $1384\pm 6^{+9}_{-0}$. We shall vary the mass of $h_1(1380)$ to estimate uncertainties.}  can be understood considering it to be a $K^* \bar K$ state, especifically,
 \begin{align}
 |h_1(1380)\rangle&=-\frac{1}{2}\Big[|K^{*+}K^-\rangle+|K^{*0}\bar K^0\rangle\nonumber\\
 &\quad+|K^- K^+\rangle+|\bar K^{*0} K^0 \rangle\Big].\label{h1}
 \end{align}
In this way, studying the $\phi (K^*\bar K)_{I=0}+\text{c.c}$ system might shed some light on the nature of $X(2370)$. Note that the interaction of $\phi$  with all the components of Eq.~(\ref{h1}) is the same; hence, we can focus on studying, for example, $\phi (K^*\bar K)_{I=0}$. To do this,  we consider the fixed center approximation (FCA) (see the review paper~\cite{MartinezTorres:2020hus}), where $\phi$ interacts with the constituents of the cluster, i.e., the $K^*$ and $\bar K$ forming $h_1(1380)$. In particular, we follow a recent version of the FCA implementing elastic unitarity~\cite{Ikeno:2025bsx} [in the present case, of $\phi$ and the cluster in Eq.~(\ref{h1})]. In this context, it is important to mention that this method, when applied to the study of the $p f_1(1285)$ interaction, produces a correlation function~\cite{Encarnacion:2026zas} in good agreement with a recent experimental measurement by the ALICE collaboration~\cite{laura}. The diagrams included in the evaluation of the scattering matrix of $\phi$ and the $K^*\bar K$ cluster can be found in Figs. 1 and 2 of Ref.~\cite{Jia:2026iqo}, where the FCA is applied to investigate a system made of mesons, as is the case here. The scattering amplitude describing the interaction between $\phi$ and the cluster of $K^*\bar K$ is then given by~\cite{Agatao:2025ckp}
 \begin{align}
 T=\frac{\tilde{t}_1+\tilde{t}_2+(2G_0-G^{(1)}_C-G^{(2)}_C)\tilde{t}_1\tilde{t}_2}{1-G^{(1)}_C\tilde{t}_1-G^{(2)}_C\tilde{t}_2-(G^2_0-G^{(1)}_CG^{(2)}_C)\tilde{t}_1\tilde{t}_2}\label{T}
 \end{align}
where
\begin{align}
\tilde{t}_1=\frac{M_C}{M_{\bar K^*}}t_1,~\tilde{t}_2=\frac{M_C}{M_{\bar K}}t_2.\label{t12}
\end{align}
In Eq.~(\ref{t12}), $M_C$ is the mass of the cluster [$h_1(1380)$] and $t_1$, $t_2$ are the scattering matrices for $\phi K^*$ and $\phi \bar K$, respectively. The normalization factors present in Eq.~(\ref{t12}) are introduced to connect the $T$-matrices of $\phi K^*$, $\phi \bar K$ with that of $\phi h_1(1380)$. In Eq.~(\ref{T}), $G_0$, $G^{(1)}_C$, $G^{(2)}_C$ stand for the $\phi$ propagator folded with the $K^*\bar K$ wave function of $h_1(1380)$ for the $\phi$ propagating from $K^*$ to $\bar K$ in the cluster ($G_0$) and from $K^*$ to $K^*$ in the coherent propagation of $\phi$ and $h_1$ as a whole ($G^{(1)}_C$), or from $\bar K$ to $\bar K$ in the coherent propagation of $\phi$ and $h_1$ ($G^{(2)}_C$). Their expressions can be found in Eqs.~(6) and (14) of Ref.~\cite{Jia:2026iqo}, with the change $D\to K^*$, $K\to \bar K$ for the cluster and the external particle $K^0\to \phi$. The arguments of $t_1$ and $t_2$ are related to the $\sqrt{s}$ value of the $\phi h_1$ system in their rest frame by means of Eq.~(3) of Ref.~\cite{Jia:2026iqo}, replacing the masses as explained above.

\section{Discussion of possible decay modes}
\begin{figure}
\centering
\includegraphics[width=0.2\textwidth]{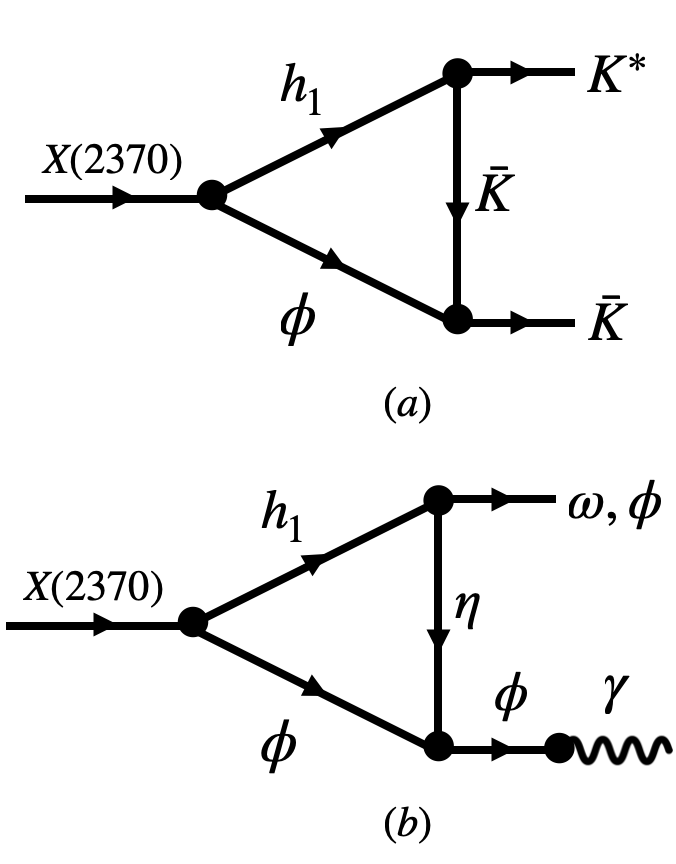}\\
\caption{Mechanism for $X(2370)$ decaying to: (a) $K^*\bar K$ and (b) $\omega\gamma$, $\phi\gamma$.}\label{dec1}
\end{figure}

In Fig.~\ref{dec1}(a), we show how, in the present picture, $X(2370)$ can in principle decay to $K^* \bar K$. As we can see in the figure, the $h_1$ in the loop can decay to $K^*\bar K$ when considering the width of $h_1$, but the lower vertex involves $\phi\bar K\to\bar K$, or, equivalently, $\bar K\to \phi\bar K$. The latter implies that the internal $\phi \bar K$ are very far off-shell. This suppresses the loop extraordinarily, making the decay amplitude very small.

In Fig.~\ref{dec1}(b), we show a mechanism for $X(2370)$ decaying to $\omega\gamma$. As shown in Table V of Ref.~\cite{Roca:2005nm}, $h_1(1380)$ (which appears with an approximate mass of 1245 MeV) couples mostly to $K^*\bar K$, but also to $\phi\eta$ and $\omega\eta$. Hence, the process depicted in Fig.~\ref{dec1}(b) for $X(2370)\to\gamma\omega$ is quite natural, but it again involves the vertex $\phi\eta\to\gamma$, which is kinematically forbidden, forcing the $\phi\eta$ particles to be far off-shell. This and the reduced $h_1\omega\eta$ coupling make the decay width related to this mechanism very small. The $h_1(1380)$ also couples to $\phi\eta$ with about the same strength as to $\omega\eta$ and hence the process of Fig.~\ref{dec1}(b) for $X(2370)\to\phi\gamma$ is equally suppressed.

\begin{figure}
\centering
\includegraphics[width=0.2\textwidth]{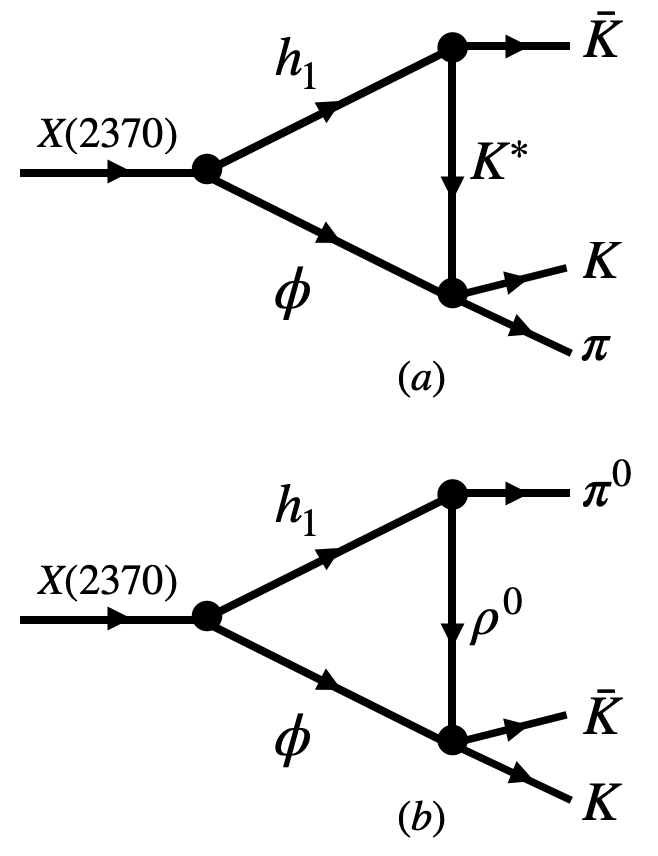}
\caption{Mechanism for $X(2370)$ decaying to: (a) $\pi K\bar K$, consistent with the $K^*\bar K$ nature of $h_1(1380)$. A triangle singularity develops in this process. (b) $\pi K\bar K$, where $\pi K$ is not obtained from the decay of $K^*(892)$.}\label{dec3}
\end{figure}

In Fig.~\ref{dec3}(a), we show the mechanism for $X(2370)$ decaying to $\pi K\bar K$, considering the $K^*\bar K$ nature of $h_1(1380)$. The $\phi K^*$ interaction is studied in Ref.~\cite{Geng:2008gx}, where the transition to $\pi K$ is considered. The process depicted in Fig.~\ref{dec3}(a) produces a triangle singularity when $M_X\sim 2433$ MeV, which corresponds to the $\phi h_1$ threshold, and the $\pi K$ invariant mass is $M_{\pi K}\sim 1912$ MeV. 

Another possible decay mechanism to produce the same final state is depicted in Fig.~\ref{dec3}(b). Once again, $h_1(1380)$ also couples to $\rho\pi$~\cite{Roca:2005nm}, and now the process is not hindered by the off-shell restrictions from off-shell propagators in the loop. Furthermore, the $\phi\rho$ interaction, studied in Ref.~\cite{Geng:2008gx}, together with other coupled channels, produces an $a_0$ resonance with quantum numbers $I^G(J^{PC})=1^-(0^{++})$. Adding a $\pi^0$ [$1^-(0^{-+})$] to the former $a_0$ match the quantum numbers of $X(2370)$. The mentioned $a_0$ has a large width, $\Gamma\simeq 150$ MeV, as a consequence of $\rho\phi\to K\bar K$, hence, the amplitude involved in Fig.~\ref{dec3}(a) for $\rho\phi\to K\bar K$ has a large strength. Interestingly, this predicted resonance, also obtained in the approach of Ref.~\cite{Du:2018gyn}, has been recently observed in BaBar~\cite{BaBar:2021fkz}, BESIII~\cite{BESIII:2022npc}, and LHCB~\cite{LHCb:2023evz}. Discussions concerning this resonance and present and future experiments around it can be seen in Ref.~\cite{Oset:2023hyt}. More interestingly, by taking a $K\bar K$ invariant mass around 1830 MeV, close to the peak of $a_0(1790)$ when considering its width, the diagram of Fig.~\ref{dec3}(a) develops a triangle singularity (TS) for a mass of $X(2370)$ of about 2400 MeV, also close considering the $X(2370)$ width, as can be seen by a direct application of Eq.~(18) of Ref.~\cite{Bayar:2016ftu}. Within our description of $X(2370)$, the enhancement produced by the TS and the resonant behavior of the $\phi\rho\to K\bar K$ amplitude optimize the mechanism of Fig.~\ref{dec3}(b) to produce the final state $K\bar K\pi^0$ from the decay of $X(2370)$.

\section{Results}
First, we must evaluate the $t_1$ and $t_2$ amplitudes describing, respectively, the $\phi K^*$ and $\phi \bar K$ interactions. In this study, we need the $\phi K^*$ system with the quantum numbers $J^P=0^+$. The latter, together with those of the $\bar K$ ($J^P=0^-$) in the cluster, gives rise to the $J^P=0^-$ quantum numbers of $X(2370)$ (the positive C-parity of this state is given by the product of the C-parity of the $\phi$, $-1$, and that of $h_1$, $-1$ too). Here, we follow the general study of the vector-vector interaction performed in Ref.~\cite{Geng:2008gx}, where amplitudes in the $J^P=0^+$, $1^+$, and $2^+$ sectors are obtained. As can be seen in Table III of Ref.~\cite{Geng:2008gx}, for $J^P=0^+$, a resonance with strangeness 1, isospin $I=1/2$ is obtained with mass $\sim 1640$ MeV and $\Gamma\sim 48$ MeV. We identify this resonance with the $K(1630)$ reported by the PDG with unknown $J^P$, mass $1629\pm 7$ MeV and $\Gamma\simeq 16^{+19}_{-18}$ MeV. In this way, we consider a Breit-Wigner form for the $\phi K^*$ amplitude in the $J^P=0^+$ sector, with the mass and width given from the PDG and the coupling from Ref.~\cite{Geng:2008gx},
\begin{align}
&t_1=\frac{g^2_{K(1630) \phi K^*}}{s_1-M^2_{K(1630)}+i M_{K(1630)} \Gamma_{K(1630)}};\nonumber\\
&g_{K(1630)\phi K^*}=-1518+i209~\text{MeV}.\label{K1630}
\end{align}

Next, we need the $\phi \bar K$ interaction. This system has been studied with a coupled-channel approach in Ref.~\cite{Roca:2005nm}, showing the dynamical generation of $K_1(1270)$ with a double-pole nature, but no signal for the generation of $K_1(1400)$ was found. For the energy region under study, the invariant mass of the $\phi \bar K$ system is closer to the energy region where $K_1(1400)$ appears; thus, it would be better to know the coupling of $K_1(1400)$ to $\phi K$, independently of the nature of the mentioned state. A study done in Ref.~\cite{Malabarba:2020grf}, where radiative decays of $K_1(1400)$ and $K_1(1270)$ are studied considering the model of Ref.~\cite{Palomar:2003rb} shows that the coupling of $K_1(1400)$ to $\phi \bar K$ is strong, with a value of $3522\pm 21$ MeV, larger than that of Eq.~(\ref{K1630}) for $K(1630)$, and also bigger than the couplings to $\phi K$ obtained in Ref.~\cite{Roca:2005nm} for the two poles of $K_1(1270)$ . Thus, we again take an amplitude
\begin{align}
&t_2=\frac{g^2_{K_1(1460)\phi K}}{s_2-M^2_{K_1(1400)}+i M_{K_1(1400)} \Gamma_{K_1(1400)}},\nonumber\\
&g_{K_1(1400)\phi\bar K}=3522\pm 21~\text{MeV}.
\end{align}
It is worth noting that after the study performed in Ref.~\cite{Malabarba:2020grf}, the LHCb collaboration has reported a strong coupling of $K_1(1400)$ to $\phi K$, allowed within the span of the $K_1$ width, with a sizeable fit fraction of about 15\%~\cite{LHCb:2021uow}. In any case, we will vary the mass and coupling to $\phi K$ of $K_1(1400)$ in a wider range to estimate uncertainties.

With this input on the $t_1$ and $t_2$ amplitudes, we determine the $T$-matrix of Eq.~(\ref{T}) and investigate the possible formation of a state below the $\phi h_1$ threshold. Before proceeding with further discussions, we should mention that the evaluation of $G_0$, $G^{(i)}_C$, and the cluster form factors involved requires the use of a regulator, $q_\text{max}$, in the loop integrals.  Based on the work of Ref.~\cite{Geng:2008gx},  $q_\text{max}=971.5$ MeV, but we vary it in the interval $900-1000$ MeV to estimate uncertainties.

Next, we determine the line shape of the $K^0_S K^0_S\pi^0$ invariant mass distribution in the process $J/\psi\to\gamma K^0_S K^0_S\pi^0$ and compare it with the experimental results, where the mass distribution of $K^0_S K^0_S\pi^0$ was shown. To do this, we can treat the $\gamma K^0_S K^0_S\pi^0$ final state as a two-body system formed by a $\gamma$ and the $K^0_S K^0_S\pi^0$ subsystem, the latter with an invariant mass $M_\text{inv}$. In this way, for a given value of $M_\text{inv}$, the decay width of $X(2370)$ to the mentioned final state can be written as
\begin{align}
\Gamma(M_\text{inv})=\frac{1}{8\pi}\frac{1}{m_{J/\psi}^2}|t|^2 p_\gamma. \label{Gamma}
\end{align}
In Eq.~(\ref{Gamma}), $t$ represents the amplitude describing the process $J/\psi\to\gamma K^0_S K^0_S\pi^0$ and $p_\gamma$ is the modulus of the linear momentum of the photon in the rest frame of $\gamma$ and the $K^0_S K^0_S\pi^0$ subsystem, i.e.,
\begin{align}
p_\gamma=\frac{\lambda^{1/2}(m^2_{J/\psi},m^2_\gamma,M^2_\text{inv})}{2m_{J/\psi}}.
\end{align}
Considering the distribution related to all kinematically allowed values of $M_\text{inv}$, and the presence of $X(2370)$ in such a distribution, to calculate the decay width for $J/\psi\to\gamma K^0_S K^0_S\pi^0$, we should fold Eq.~(\ref{Gamma}) with the spectral function associated with $X(2370)$ (whose mass and width are $M_X$ and $\Gamma_X$, respectively), i.e.,
\begin{align}
\Gamma=\int dM^2_{\text{inv}}\left(-\frac{1}{\pi}\right)\text{Im}\Big[\frac{1}{M^2_\text{inv}-M^2_{X}+i M_X\Gamma_X}\Big]\Gamma(M_\text{inv}).\label{Gammaf}
\end{align}
The imaginary part present in Eq.~(\ref{Gammaf}) can be related to the $T$-matrix obtained from Eq.~(\ref{T}), considering a Breit-Wigner form for energies close to the observed peak position, and which is associated with $X(2370)$. Indeed, we can write
\begin{align}
T(\sqrt{s})=\frac{g^2_{X \phi h_1}}{s-M^2_X+iM_X\Gamma_X}, \label{TBW}
\end{align}
where $g_{X\phi h_1}$ is the coupling of the generated $X(2370)$ to $\phi h_1(1380)$. In the present context of calculating the decay width $J/\psi\to\gamma K^0_S K^0_S\pi^0$, the $\sqrt{s}$ in Eq.~(\ref{TBW}) would precisely correspond to $M_\text{inv}$, such that
\begin{align}
|\text{Im}[T(M_\text{inv})]|\propto -\text{Im}\Big[\frac{1}{M^2_\text{inv}-M^2_{X}+i M_X\Gamma_X}\Big].
\end{align}
Assuming that for energies close to the mass of $X(2370)$ the amplitude $t$ present in Eq.~(\ref{Gamma}) has a very mild energy dependence compared to that of Eq.~(\ref{TBW}), we can write
\begin{align}
\frac{d\Gamma}{dM_\text{inv}}\propto |\text{Im}[T(M_\text{inv})]|\frac{M_\text{inv}}{m_{J/\psi}}(m^2_{J/\psi}-M^2_\text{inv}).\label{dGammadMinv}
\end{align}
To obtain Eq.~(\ref{dGammadMinv}) we have explicitly substituted the $p_\gamma$ in Eq.~(\ref{Gamma}) in terms of $M_\text{inv}$ and $m_{J/\psi}$.

\begin{figure}
\centering
\includegraphics[width=0.4\textwidth]{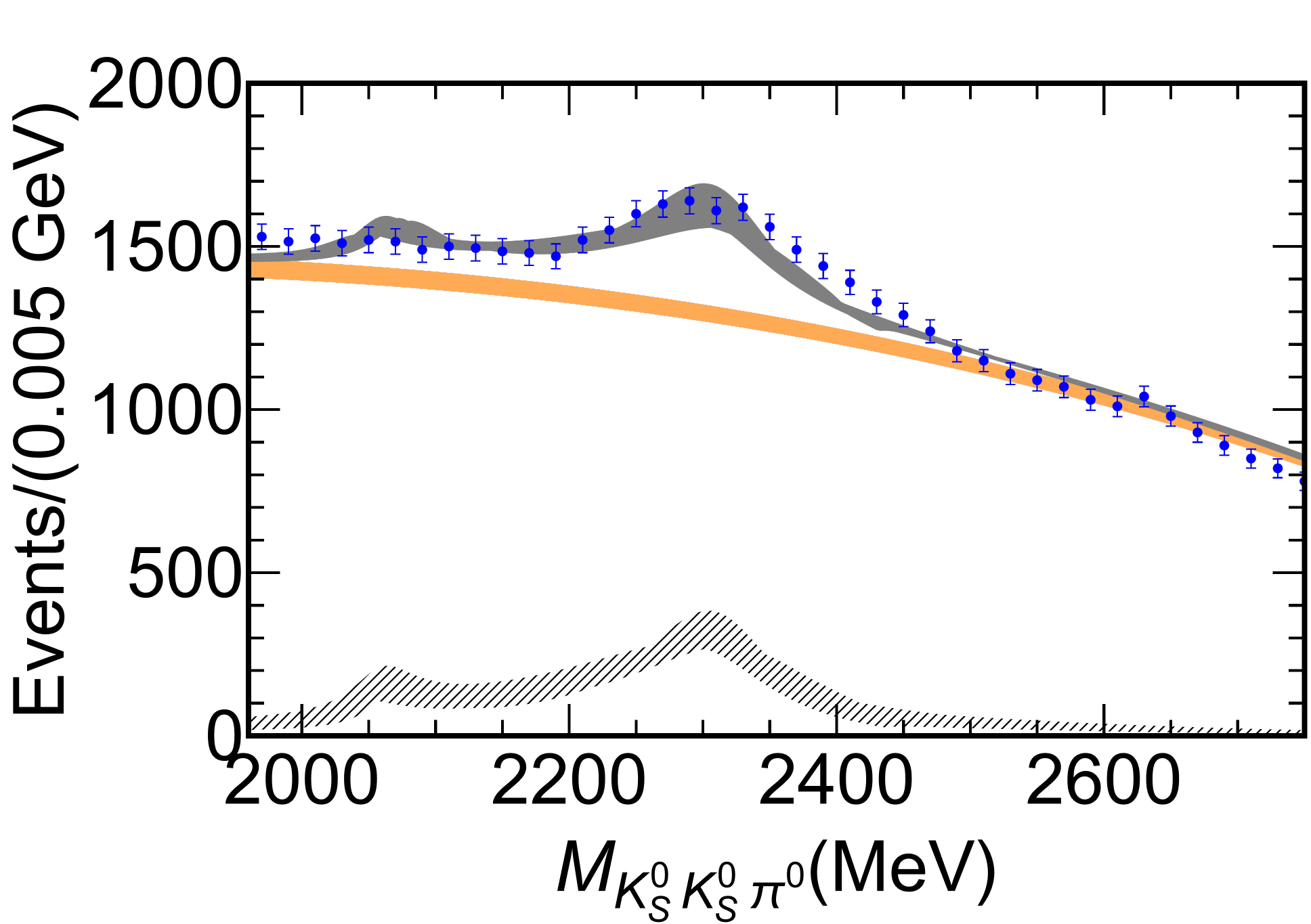}
\caption{Invariant mass distribution of the $K^0_SK^0_S\pi^0$ system in the $J/\psi\to\gamma K^0_SK^0_S\pi^0$ decay. The experimental data (dots) are taken from Ref.~\cite{BESIII:2026rzt}. The shaded regions represent uncertainties obtained when generating random numbers for the masses of $K_1(1400)$, $K(1630)$, $h_1(1380)$, the widths of $K_1(1400)$, $K(1630)$, and the coupling of $K_1(1400)$ to $\phi K$. The lighter shaded region represents the non-resonant contribution obtained from a third-order polynomial, as in Ref.~\cite{BESIII:2026rzt}. The darker shaded region includes the resonant contribution (shown as a hatched region).}\label{Minv}
\end{figure}

In Fig.~\ref{Minv}, we show the results obtained for the invariant mass distribution of the $K^0_SK^0_S\pi^0$ system in the process $J/\psi\to \gamma K^0_SK^0_S\pi^0$. Following Ref.~\cite{BESIII:2026rzt}, we focus on the energy region $M_\text{inv}\sim 2-2.7$ GeV and consider a third-order polynomial in $M_{\text{inv}}$ to describe the non-resonant signal observed. We then consider an incoherent sum of this polynomial and the contribution obtained from Eq.~(\ref{dGammadMinv}). The coefficients appearing in the mentioned polynomial and the proportionality constant in Eq.~(\ref{dGammadMinv}), which acts as a relative weight between non-resonant and resonant signals, are fitted to the data. To estimate uncertainties, we generate random numbers for the masses of $h_1(1380)$, $K_1(1400)$, $K(1630)$, as well as the widths of $K_1(1400)$, and $K(1630)$, based on the data collected by the PDG: $M_{K_1(1400)}\sim 1400-1500$ MeV, $\Gamma_{K_1(1400)}\sim 100-200$ MeV, $M_{h_1(1380)}\sim 1360-1420$ MeV, $M_{K(1630)}\sim 1620-1640$ MeV, $\Gamma_{K(1630)}\sim 16-40$ MeV. In view of the mass uncertainty of $K_1(1400)$, we have varied as well $g_{K_1(1400) \phi K}$. In particular, we have considered $g_{K_1(1400)\phi K}\sim 2000-4000$ MeV. We find that a satisfactory description of the data is obtained for $M_{K_1(1400)}\sim 1460-1480$ MeV, $\Gamma_{K_1}\sim 130-220$ MeV, $g_{K_1(1400)\phi K}\sim 2500-3600$, $M_{h_1}\sim 1380-1412$ MeV, $M_{K(1630)}\sim 1611-1647$ MeV, $\Gamma_{K(1630)}\sim 36-67$ MeV, as shown by the (darker) shadded region in Fig.~\ref{Minv}. The hatched region in Fig.~\ref{Minv} represents the result obtained from Eq.~(\ref{dGammadMinv}), which is proportional to the $|T|^2$ describing the $\phi h_1(1380)$ system [and obtained using Eq.~(\ref{T})]. Two peaks can be observed: the first one, which appears for $M_\text{inv}\sim 2050$ MeV, is a manifestation of the $K(1630)$ generated in the $\phi K^*$ system: for an energy of the $\phi h_1$ system around 2050 MeV, the invariant mass of the $\phi K^*$ subsystem is in the mass region of $K(1630)$, producing a peak in the three-body $T$-matrix at that value of $\sqrt{s}$. In fact, this peak position can be observed already within the impulse approximation, i.e., ignoring all rescattering contributions. Similarly, another peak is observed within the impulse approximation for a value of $\sqrt{s}$ at which the invariant mass of the $\phi K$ system is in the $K_1(1400)$ region. The second peak observed in the hatched region, however, appears about 30 MeV below the previously mentioned energy, indicating that the considered three-body dynamics has bound the system and generated a genuine state with a mass of $\sim 2300$ MeV.  Considering the uncertainties, we find the mass and width of the state to be $M_X=2316\pm 10$ MeV and $\Gamma_X=163\pm 31$ MeV, in line with the values found for $X(2370)$.

\section{Conclusions}
We have presented a description of the properties of $X(2370)$ in which the state is generated from the three-body dynamics involved in the $\phi h_ 1(1380)$ system. The mass and width obtained are $2316\pm10$ MeV and $163\pm 31$ MeV, respectively. We also showed that the quantum numbers obtained from $\phi$ interacting with $h_1(1380)$ match perfectly with those of $X(2370)$. The amplitudes needed for the evaluation of the $\phi h_ 1(1380)$ amplitude were borrowed from previous works reproducing observables of different reactions and then, within uncertainties that we evaluate, the obtained results are parameter-free. Equally relevant is the fact that we showed that the modes to $\gamma \omega$ and $\gamma \phi$, together with the decay to $K^* \bar K$, whose suppression is invoked in Ref.~\cite{BESIII:2026mvn} as indicative of the glueball nature of $X(2370)$, appear naturally suppressed in our approach. Conversely, the decay mode to $K\bar K \pi$ of non $K^* \bar K$ nature, shown to be relevant in Ref.~\cite{BESIII:2026mvn}, is largely enhanced in our picture due to the combination of a triangle singularity and the presence of the $K(1630)$ resonance coupling strongly to $\phi \rho$ and decaying to $K \bar K$. Given the TS nature of this decay, it would be most instructive to look at the  $K \bar K$ and $\pi K$ mass distributions, not reported in the experimental paper (see Ref.~\cite{BESIII:2026mvn}), where large contributions around $M_{K\bar K}$ of the order of 1830 MeV, and $M_{\pi K}$ of the order of 1912 MeV, are predicted. 

\section{Acknowledgements}
This work is supported by the Grants PID2023-147458NB-C21 and CEX2023-001292-S funded by MICIU/AEI/10.13039/501100011033 and by ERDF/EU, as well as of the PROMETEO program Grant CIPROM/2023/59 funded by Generalitat Valenciana 10.13039/501100003359. We thank CNPq (K.P.K: Grants No. 407437/ 2023-1 and No. 306461/2023-4; A.M.T: Grant No. 304510/2023-8), FAPESP (A. M. T.: Grant number 2025/20068-6) for their support.

\bibliographystyle{unsrt}
\bibliography{refs}

\end{document}